\documentclass[nonacm]{acmart}

\usepackage{algorithm}
\usepackage[noend]{algorithmic}

\newcommand{\addcsch}[1]{{#1}}
\newcommand{\cschreplace}[2]{#2}

\usepackage{multirow}
\usepackage{numprint}

\newtheorem{mytheorem}{Theorem}

\DeclareRobustCommand{\frcshape}{\fontfamily{frc}\selectfont}
\DeclareTextFontCommand{\textfrc}{\frcshape}
\newcommand{\mathfrc}[1]{\mbox{\frcshape #1}}

\def\MdR{\ensuremath{\mathbb{R}}}

\newcommand{\set}[1]{\left\{ #1\right\}}
\newcommand{\sodass}{\,:\,}
\newcommand{\setGilt}[2]{\left\{ #1\sodass #2\right\}}

\newcommand{\Is}       {:=}

\def\comment#1{}

\def\withcomments{
	\newcounter{mycommentcounter}
	\def\comment##1{\refstepcounter{mycommentcounter}%
		\ifhmode%
		\unskip%
		{\dimen1=\baselineskip \divide\dimen1 by 2 %
			\raise\dimen1\llap{\tiny\bfseries \textcolor{red}{-\themycommentcounter-}}}\fi%
		\marginpar[{\renewcommand{\baselinestretch}{0.8}%
			\hspace*{3em}\begin{minipage}{5em}\footnotesize [\themycommentcounter]: \raggedright ##1\end{minipage}}]{\renewcommand{\baselinestretch}{0.8}%
			\begin{minipage}{5em}\footnotesize [\themycommentcounter]: \raggedright ##1\end{minipage}}}
}

\begin{document}
\title{Local Motif Clustering via (Hyper)Graph Partitioning}

\author{Adil Chhabra}
\affiliation{%
	\institution{Heidelberg University}
	\country{Germany}
}
\email{adil.chhabra@stud.uni-heidelberg.de}

\author{Marcelo Fonseca Faraj}
\affiliation{%
	\institution{Heidelberg University}
	\country{Germany}
}
\email{marcelofaraj@informatik.uni-heidelberg.de}

\author{Christian Schulz}
\affiliation{%
	\institution{Heidelberg University}
	\country{Germany}
}
\email{christian.schulz@informatik.uni-heidelberg.de}

\begin{abstract}
A widely-used operation on graphs is local clustering, i.e., extracting a well-characterized community around a seed node without the need to process the whole graph.  Recently local motif clustering has been proposed: it looks for a local cluster based on the distribution of motifs.  Since this local clustering perspective is relatively new, most approaches proposed for it are extensions of statistical and numerical methods previously used for edge-based local clustering, while the available combinatorial approaches are still few and relatively simple.  In this work, we build a hypergraph and a graph model which both represent the motif-distribution around the seed node. We solve these models using  sophisticated combinatorial algorithms designed for (hyper)graph partitioning.  In extensive experiments with the triangle motif, we observe that our algorithm computes communities with a motif conductance value being one third on average in comparison against the communities computed by the state-of-the-art tool MAPPR while being $6.3$ times faster on average. 
\end{abstract}

\maketitle

\section{Introduction}
\label{sec:introduction}

Graphs are a powerful mathematical abstraction which are used to represent complex phenomena such as data dependency, social networks, web links, email interactions, and so forth.
With the massive growth of the data generated on a daily basis, many real-world graphs become more and more massive, hence processing them becomes an increasing challenge.
In particular, many applications do not need to process the entire network but only a tiny and localized portion of~it, which is the case for community-detection on Web~\cite{epasto2014reduce} and social~\cite{jeub2015think} networks as well as structure-discovery in bioinformatics~\cite{voevodski2009spectral} networks among others.
Those real-world applications are usually preceded by or modeled as a \emph{local clustering}. 
Given a network, the local clustering problem consists of identifying a single \emph{well-characterized} cluster which contains a given seed node or most of a given set of seed nodes.
Well-characterized here means that the computed cluster ideally contains many internal edges and few external edges.
More specifically, the quality of a community can be quantified by metrics such as \emph{conductance}~\cite{kannan2004clusterings}.
Since minimizing conductance is NP-hard~\cite{wagner1993between}, approximative and heuristic approaches are used in practice.
In light of the nature of this problem and its scalability requirement, these approaches ideally require time and memory dependent only on the size of the returned cluster.

The local clustering problem has been studied both theoretically~\cite{andersen2006local} and experimentally~\cite{leskovec2009community}, and has been solved using a wide variety of methods, including statistical~\cite{chung2013solving,kloster2014heat}, numerical~\cite{li2015uncovering,mahoney2012local}, and combinatorial~\cite{orecchia2014flow,fountoulakis2020flowbased} approaches.
Most works on local clustering evaluate the quality of a local community exclusively based on the given distribution of edges.
Nevertheless, some novel approaches~\cite{yin2017local,zhang2019local,meng2019local,murali2020online} go in a different direction by finding local communities based on the distribution of higher-order structures which are known as motifs.
These works provide experimental evidence that this approach, which can be called \emph{local motif clustering}, is promising on detecting high-quality local communities.
Nevertheless, since this local clustering perspective is relatively new, most approaches proposed for it are extensions of statistical and numerical methods previously proposed for edge-based local clustering or are very simple combinatorial algorithms.

In this work, we employ sophisticated combinatorial algorithms as a tool to solve the local motif clustering problem.
We propose two algorithms, one uses a graph model and the other one uses a hypergraph model. 
Our algorithm starts by building a (hyper)graph model which represents the motif-distribution around the seed node on the original graph.
While the graph model is exact for motifs of size at most three, the hypergraph model works for arbitrary motifs and is designed such that an optimal solution in the (hyper)graph model minimizes the motif conductance in the original network.
The (hyper)graph model is then partitioned using a powerful multi-level hypergraph or graph partitioner in order to directly minimize the motif conductance of the corresponding partition in the original graph.
Extensive experiments evaluate the trade-offs between the two different models. Moreover, when using the graph model for triangle motifs, our algorithm computes communities that have on average one third of the motif conductance value than communities computed by MAPPR while being $6.3$ times faster on average and removing the necessity of a preprocessing motif-enumeration on the whole network.

\section{Preliminaries}
\label{sec:preliminaries}

\label{subsec:basic_concepts}

Let $G=(V=\{0,\ldots, n-1\},E)$ be an \emph{undirected graph} with no multiple or self edges allowed, such that $n = |V|$ and $m = |E|$.
Let $c: V \to \MdR_{\geq 0}$ be a node-weight function, and let $\omega: E \to \MdR_{>0}$ be an edge-weight function.
We generalize $c$ and $\omega$ functions to sets, such that $c(V') = \sum_{v\in V'}c(v)$ and $\omega(E') = \sum_{e\in E'}\omega(e)$.
Let $N(v) = \setGilt{u}{\set{v,u}\in E}$ be the \emph{open neighborhood} of $v$, and let $N[v]=N(v) \cup \{v\}$ be the \emph{closed neighborhood} of $v$.
We generalize the notations $N(.)$ and $N[.]$ to sets, such that $N(V') = \cup_{v\in V'}N(v)$ and $N[V'] = \cup_{v\in V'}N[v]$.
A graph $G'=(V', E')$ is said to be a \emph{subgraph} of $G=(V, E)$ if $V' \subseteq V$ and $E' \subseteq E \cap (V' \times V')$. 
When $E' = E \cap (V' \times V')$, $G'$ is the subgraph \emph{induced} in $G$ by $V'$.
Let $\overline{V'} = V \setminus V'$ be the \emph{complement} of a set $V' \subseteq V$ of nodes. 
Let a \emph{motif} $\mu$ be a connected graph.
\emph{Enumerating} the motifs $\mu$ in a graph $G$ consists building the collection $M$ of all occurrences of $\mu$ as a subgraph of $G$.
Let $d(v)$ be the degree of node $v$ and $\Delta$ be the maximum degree of $G$.
Let $d_\omega(v)$ be the weighted degree of a node $v$ and $\Delta_\omega$ be the maximum weighted degree of $G$.
Let $d_\mu(v)$ be the number of motifs $\mu \in M$ which contain $v$.
We generalize the notations $d(.)$, $d_\omega(.)$, and $d_\mu(.)$ to sets, such that $d(V') = \sum_{v\in V'}d(v)$, $d_\omega(V') = \sum_{v\in V'}d_\omega(v)$, and $d_\mu(V') = \sum_{v\in V'}d_\mu(v)$.
Let a \emph{spanning forest} of $G$ be an acyclic subgraph of $G$ containing all its nodes.
Let the \emph{arboricity} of $G$ be the minimum amount of spanning forests of $G$ necessary to cover all its edges.

In the \emph{local graph clustering} problem, a graph $G=(V,E)$ and a seed node $u \in V$ are taken as input and the goal is to detect a \emph{well-characterized cluster} (or \emph{community}) $C \subset V$ containing~$u$.
A high-quality cluster $C$ usually contains nodes that are densely connected to one another and sparsely connected to $\overline{C}$.
There are many functions to quantify the quality of a cluster, such as \emph{modularity}~\cite{brandes2007modularity} and \emph{conductance}~\cite{kannan2004clusterings}.
The conductance metric is defined as $\phi(C)=|E'|/\min(d(C),d(\overline{C}))$, where $E'=E \cap (C \times \overline{C})$ is the set of edges shared by a cluster $C$ and its complement.
\emph{Local motif graph clustering} is a generalization of local graph clustering where a motif $\mu$ is taken as an additional input and the computed cluster optimizes a clustering metric based on $\mu$.
In particular, the \emph{motif conductance} $\phi_\mu(C)$ of a cluster $C$ is defined by 
\citet{benson2016higher} as a generalization of the conductance in the following way:
$\phi_\mu(C)=|M'|/min(d_\mu(C),d_\mu(\overline{C}))$, where $M'$ are all the motifs $\mu$ which contain at least one node in $C$ and one node in $\overline{C}$.
Note that, if the motif under consideration is simply an edge, then $|M'|$ is the edge cut and $\phi_\mu(C)=\phi(C)$.

Let $H=(\mathcal{V}=\{0,\ldots, \mathfrc{n}-1\},\mathcal{E})$ be an \emph{undirected hypergraph} with no multiple or self hyperedges allowed, with $\mathfrc{n} = |\mathcal{V}|$ \emph{nodes} and $\mathfrc{m} = |\mathcal{E}|$ \emph{hyperedges} (or \emph{nets}).
A net is defined as a subset of $\mathcal{V}$.
The nodes that compose a net are called \emph{pins}.
Let $\mathfrc{c}: \mathcal{V} \to \MdR_{\geq 0}$ be a node-weight function, and let $\mathfrc{w}: \mathcal{E} \to \MdR_{>0}$ be a net-weight function.
We generalize $\mathfrc{c}$ and $\mathfrc{w}$ functions to sets, such that $\mathfrc{c}(\mathcal{V}') = \sum_{v\in \mathcal{V}'}\mathfrc{c}(v)$ and $\mathfrc{w}(\mathcal{E}') = \sum_{e\in \mathcal{E}'}\mathfrc{w}(e)$.
A node $v\in\mathcal{V}$ is \emph{incident} to a net $e\in\mathcal{E}$ if $v \in e$.
Two nodes are \emph{adjacent} if they are incident to a same net.
Let the number of pins~$|e|$ in a net~$e$ be the \emph{size} of~$e$.
We define the \emph{contraction} operator as $\big/$ such that $H \big/ \mathcal{E}^\prime$, with $\mathcal{E}^\prime \subseteq \mathcal{E}$, is the hypergraph obtained by contracting the nodes from~$\mathcal{E}^\prime$on~$H$.
This contraction consists of substituting the nodes in $\mathcal{E}^\prime$ by a single representative node $x$, removing nets totally contained in $\mathcal{E}^\prime$, and substituting all the pins in $\mathcal{E}^\prime$ by a single pin $x$ in each of the remaining nets.

A \emph{$k$-way partition} of a (hyper)graph $H$ is a partition of its vertex set into $k$ \emph{blocks} $\{V_1, \dots, V_k\}$
such that $\bigcup_{i=1}^k V_i = V$, $V_i \neq \emptyset $ for $1 \leq i \leq k$, and $V_i \cap V_j = \emptyset$ for $i \neq j$.
We call a $k$-way partition \emph{$\varepsilon$-balanced} if each block $V_i$ satisfies the \emph{balance constraint}:
$c(V_i) \leq L_{\max} := (1+\varepsilon)\lceil \frac{c(V)}{k} \rceil$ for some parameter $\mathrm{\varepsilon}$. %
In the graph case, typically the edge cut is minimized, i.e. 
the total weight of the edges crossing blocks, i.e., $\sum_{i<j}\omega(E_{ij})$, where $E_{ij}\Is\setGilt{\set{u,v}\in E}{u\in V_i,v\in V_j}$. 
In the hypergraph case, one metric that is often minimized is the \emph{cut-net} which consists of the total weight of the nets crossing blocks, i.e., $\sum_{e \in \mathcal{E}^\prime}\mathfrc{w}(\mathcal{E}^\prime)$, in which $\mathcal{E}^\prime \Is $ $\big\{e \in \mathcal{E} : \exists i,j \mid e \cap \mathcal{V}_i \neq \emptyset, e \cap \mathcal{V}_j \neq \emptyset , i\neq j\}$.

\subsection{Related Work}
\label{subsec:related_work}

Many works partition all the nodes of a graph into clusters based on motifs, such as~\cite{benson2015tensor, yin2017local,klymko2014using,prvzulj2007biological,tsourakakis2017scalable}. 
Similarly to these works, we deal here with motif-based clustering, but our scope is local graph clustering around a seed node.
Several other works propose local clustering algorithms, such as~\cite{kloster2014heat,li2015uncovering,mahoney2012local,cui2014local,sozio2010community}, however they do not optimize for motif-based metrics, but rather for metrics based on edges such as conductance and modularity.
Here, we focus on contributions directly related to the scope of this work.

\citet{rohe2013blessing} propose a local clustering algorithm based on triangle motifs.
Their algorithm starts with a cluster containing only the seed node, and iteratively grows this cluster.
Particularly, the algorithm greedily inserts nodes contained in at least a predefined amount of cut triangles.
\citet{huang2014querying} recover local communities containing a seed node in online and dynamic setups based on higher-order graph structures named Trusses~\cite{cohen2008trusses}.
They define the $k$-truss of a graph as its largest subgraph whose edges are all contained in at least $(k-2)$ triangle motifs, hence it is a graph structure based on triangle-frequencies.
The authors use indexes to search for $k$-truss communities in time proportional to the size of the recovered~community. 

\citet{yin2017local}~propose MAPPR, a local motif clustering algorithm based on the Approximate Personalized PageRank (APPR) method.
MAPPR starts with a preprocessing phase where it enumerates the motif of interest in the whole input graph and builds a weighted graph~$W$.
The edges in this weighted graph only exist between nodes contained in at least one motif such that their edge weight is equal to the amount of motif occurrences containing their two endpoints.
Afterward, a local community is found on the generated graph using an adapted version of APPR.
MAPPR is able to extract local communities from directed input graphs, which is not possible for APPR alone.

\citet{zhang2019local} propose the algorithm LCD-Motif to solve the local motif clustering problem using a variant of the spectral approach.
LCD-Motif has two main differences in comparison to the traditional spectral motif clustering method.
First, it avoids computing the singular vectors by performing random walks to find potential members of the searched cluster.
They compute the span of a few dimensions of vectors after some random walks to use it as the approximate local motif spectra.
Second, the algorithm does not use $k$-means for clustering, instead it looks for the minimum 0-norm vector in the above mentioned span, such that the seed nodes are contained in its support vector.

\citet{meng2019local} propose FuzLhocd, a local motif clustering algorithm based on fuzzy arithmetic.
Its fuzzy functions aim at optimizing a new quality metric proposed by them, which is an adaptation of modularity.  
Given seed node, FuzLhocd starts by detecting probable core nodes of the searched local community based on fuzzy membership.
Next, the algorithm expands the found core nodes based on another fuzzy membership to obtain an initial cluster.

\citet{zhou2021high} propose the algorithm HOSPLOC for local motif clustering. 
HOSPLOC starts by approximately computing the distribution vector with a motif-based random walk and then truncates all small vector entries to 0 to localize the computation.
Afterward, it applies a vector-based partitioning method~\cite{spielman2013local} on the distribution vector.
Differently than their approach, our algorithm is combinatorial and has no theoretical guarantees but takes advantage of sophisticated (hyper)graph partitioning algorithms.

\citet{shang2022local} propose HSEI, a local clustering algorithm based on motif and edge information.
The algorithm starts with a cluster containing only the seed node.
Then, it inserts into the cluster a node from the neighborhood of seed selected based on motif degree. 
This cluster is grown based on a motif-based extension of the modularity function.

\section{Local Motif Graph Clustering}
\label{sec:Local Motif Graph Clustering}

We now present our overall clustering strategy, then we discuss each of its algorithmic components.

\subsection{Overall Strategy}
\label{subsec:Overall Strategy}

\begin{figure*}[t]
	\centering
	\includegraphics[width=.8\linewidth]{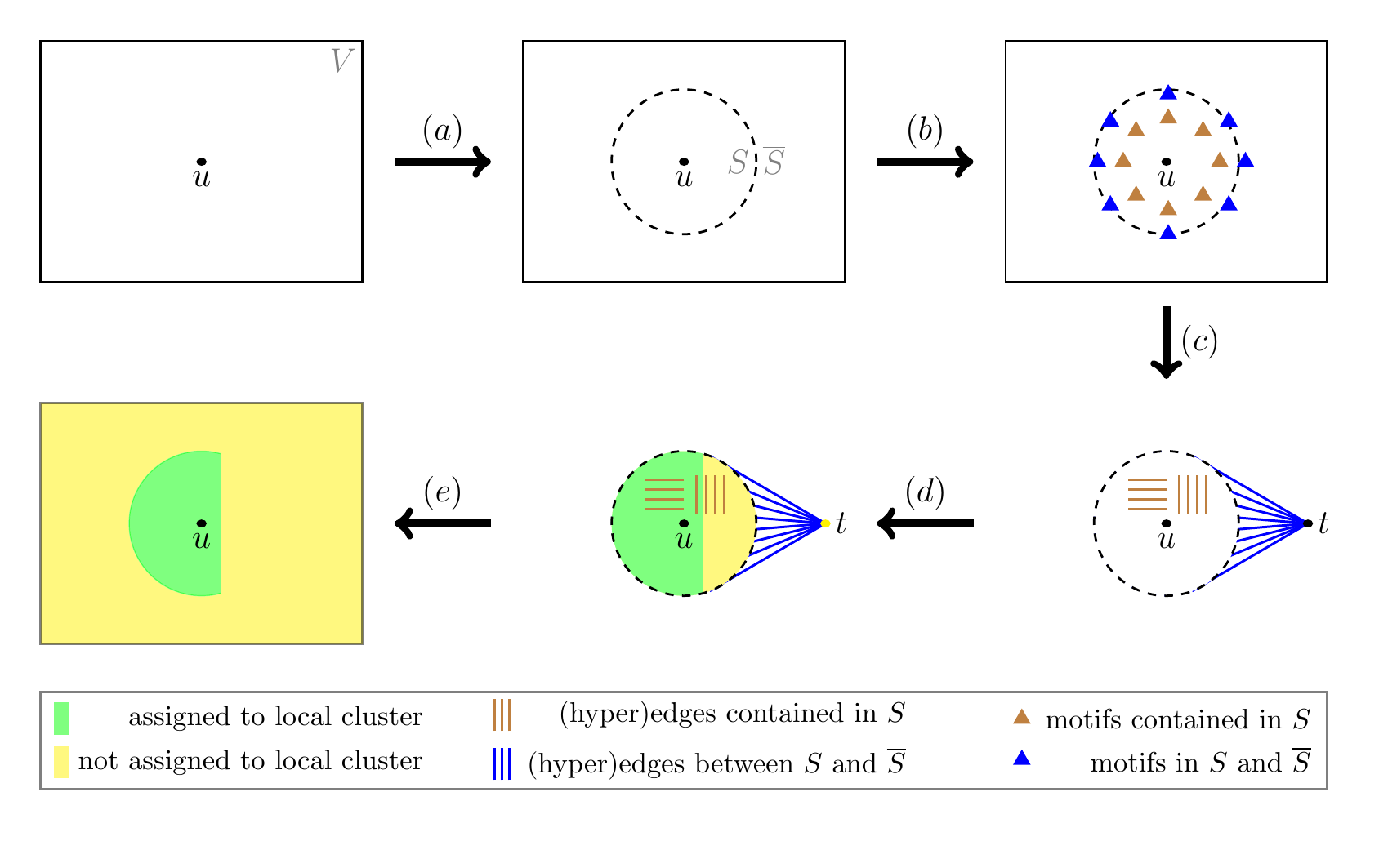}
	\caption{Illustration of the phases of our algorithm. (a)~Given a seed node $u$ and a graph $G$, a ball $S$ around $u$ is selected. (b)~Motif occurrences of $\mu$ with at least a node in $S$ are enumerated. (c)~The (hyper)graph model $H_\mu$ is built by converting motifs into (hyper)edges and contracting $\overline{S}$ into a node~$t$. (d)~The model is partitioned into two blocks using a multi-level (hyper)graph partitioner. (e)~The partition of $H_\mu$ is converted in a local cluster around the seed node in $G$.}
	\label{fig:overall_algorithm}
\end{figure*}

Given a graph $G=(V,E)$, a seed node $u$, and a motif $\mu$, our strategy for local clustering is based on four consecutive phases.
First, we select a set $S \subseteq V$ containing~$u$ and close-by nodes.
From now on, we refer to this set $S$ as a \emph{ball around} $u$.
Second, we enumerate the collection $M$ of occurrences of the motif $\mu$ which contain at least one node in~$S$.
Next, we build a graph or a hypergraph model $H_\mu$ depending on the configuration of the algorithm.
In particular, we design $H_\mu$ in such a way that the motif-conductance metric in $G$ can be computed directly in $H_\mu$.
Then, we partition this model into two blocks using a high-quality (hyper)graph partitioning algorithm.
The obtained partition of $H_\mu$ is directly translated back to $G$ as a local cluster around the seed node.
Figure~\ref{fig:overall_algorithm} provides a comprehensive illustration of the consecutive phases of our algorithm.
Note that (hyper)graph partitioning algorithms do not optimize for traditional clustering objectives such as conductance.
Instead, they aim at minimizing the edge-cut (resp. cut-net) value while respecting a hard balancing constraint.
To improve for the correct objective, we repeat the partitioning phase $\beta$ times with different imbalance constraints and pick the clustering with best motif conductance. 
Especially for the graph-based version of our model $H_\mu$, we subsequently run a special label propagation for each of these $\beta$ iterations in order to increase the chances of reaching a local minimum motif conductance.
Moreover, the first three phases of our strategy are repeated $\alpha$ times with different balls around the seed node in order to better explore the vicinity of the seed node in the original graph.
Our overall strategy including the mentioned repetitions is outlined in Algorithm~\ref{alg:overall_strategy}.

\begin{algorithm}[b!]
	\caption{Local Motif Graph Clustering}
	\label{alg:overall_strategy}
	\hspace*{-8.1cm} \textbf{Input} graph $G=(V,E)$; seed node $u \in V$; motif $\mu$ \\
	\hspace*{-7.35cm} \textbf{Output} cluster $C^* \subseteq V$ 
	\begin{algorithmic}[1]  
		\STATE $C^* \leftarrow \emptyset$
		\FOR{$i=1,\ldots,\alpha$}
		\STATE Select ball $S$ around $u$
		\STATE $M \leftarrow$ Enumerate motifs in $S$
		\STATE Build (hyper)graph model $H_\mu$ based on $S$ and $M$%
		\FOR{$j=1,\ldots,\beta$}
		\STATE Partition model $H_\mu$ into $(C, \overline{C})$, where $u \in C$
		\IF{$C^* = \emptyset \lor \phi_\mu(C) <  \phi_\mu(C^*)$}
		\STATE $C^* \leftarrow C$
		\ENDIF
		\ENDFOR
		\ENDFOR
		\STATE Convert $C^*$ into a local motif cluster in $G$
	\end{algorithmic}
\end{algorithm}

\subsection{Ball around Seed Node}
\label{subsec:Ball around Seed}

Our approach to select $S$ is a fixed-depth breadth-first search (BFS) rooted on $u$.
More specifically, we compute the first $\ell$ layers of the BFS tree rooted on $u$, then we include all its nodes in~$S$.
For each of the $\alpha$ repetitions of our overall algorithm, we use different amounts $\ell$ of layers for a better algorithm exploration. 
Two exceptional cases are handled by our algorithm, namely a ball~$S$ that is either too small or disconnected from $\overline{S}$.
We avoid the first exceptional case by ensuring that~$S$ contains $100$ or more nodes in at least one repetition of our overall algorithm.
More specifically, in case this condition is not automatically met, then we accomplish it in the last repetition by growing additional layers in our partial BFS tree while it contains fewer than $100$ nodes.
The number $100$ is based on the findings of 
\citet{leskovec2009community}, which show that most well characterized communities from real-world graphs have a relatively small size, in the order of magnitude of $100$ nodes.
If the second exceptional case happens, it means that the whole BFS tree rooted on the seed node has at most $\ell$ layers.
In this case, we simply stop the algorithm and return the entire ball $S$, which corresponds to an optimal community with motif conductance $0$ provided that there is at least one motif in $S$.

The approach described above makes sure that there is a reasonable chance that a well characterized community containing $u$ is contained in $S$, since it has at least $100$ nodes~\cite{leskovec2009community} which are all very close to $u$.
This likelihood is further increased due to the multiple repetitions of our overall algorithm using balls $S$ of different sizes.
Our BFS approach to select $S$ can be executed in time linear on the subgraph induced in~$G$ by the closed neighborhood $N[S]$ of $S$.
After selecting~$S$, the further phases of our algorithm do not deal with the whole graph, but exclusively with~$S$, its edges, and its motif occurrences.
As a consequence, our algorithm operates on a much smaller problem dimension than the size of input graph~$G$, hence its running time corresponds to the same smaller problem dimension.
The number~$\alpha$ of repetitions as well as the amount $\ell$ of layers used in each repetition are tuning~parameters.

\subsection{Motif Enumeration}

We now describe and discuss the motif-enumeration phase of our algorithm.
We optimally solve it for the triangle motif in time roughly linear on the size of the subgraph induced in~$G$ by the closed neighborhood $N[S]$ of $S$.
Moreover, we show that there are good heuristics approaches to enumerate higher-order motifs~efficiently.
The general problem of finding out if a given motif is a subgraph of some graph is NP-hard~\cite{read1977graph}, hence the enumeration of all such motifs is also NP-hard.
Nevertheless, some simpler motifs can be enumerated in polynomial time, which is the case for the triangle motif~\cite{ortmann2014triangle}.
Triangles, which can be defined as cycles or cliques of length three, have a wide variety of relevant applications on network analysis and clustering~\cite{holme2002growing,batagelj2007short,prat2012shaping}.
Without loss of generality, we specifically focus on the triangle motif within our algorithm. 
Nevertheless, note that many other small motifs can also be polynomially enumerated, such as small (directed and undirected) paths and cycles.
Moreover, our overall algorithm can also be adapted for more arbitrary motifs if we relax the optimality of the enumeration, which can be done using efficient heuristics such as the one proposed by 
\citet{kimmig2017shared}.
A simple and exact algorithm for triangle enumeration was proposed by 
\citet{chiba1985arboricity}.
Roughly speaking, this algorithm works by intersecting the neighborhoods of adjacent nodes.
For each node $v$, the algorithm starts by marking its neighbors with degree smaller than or equal to its own degree.
For each of these specific neighbors of $v$, it then scans its neighborhood and enumerates new triangles as soon as marked nodes are found.
The running time of this algorithm is $O(ma) = O(m^{\frac{3}{2}})$, where $a$ is the arboricity of the graph.
For the motif-enumeration phase of our algorithm, we apply the algorithm of 
\citet{chiba1985arboricity} only on the subgraph induced in $G$ by $N[S]$.
This is enough to find all triangles containing at least one node in $S$, as exemplified by transformation~(a) in figures~\ref{fig:model_construction}~and~\ref{fig:model_construction_graph}. 
Assuming a constant-bounded arboricity, the overall cost of our motif-enumeration phase for triangles is $O\big(|N[S] \times N[S])\cap E|\big)$.

\subsection{Hypergraph Model}
\label{subsec:Hypergraph model}

In this section, we conceptually describe the hypergraph version of our model~$H_\mu$ and explain how to build it.
We show that it can be constructed in time linear on the amount of nodes in~$S$ and motifs in~$M$.
We also discuss advantages and limitations of our hypergraph model approach.

\begin{figure*}[t!]
	\centering
	\includegraphics[width=.8\linewidth]{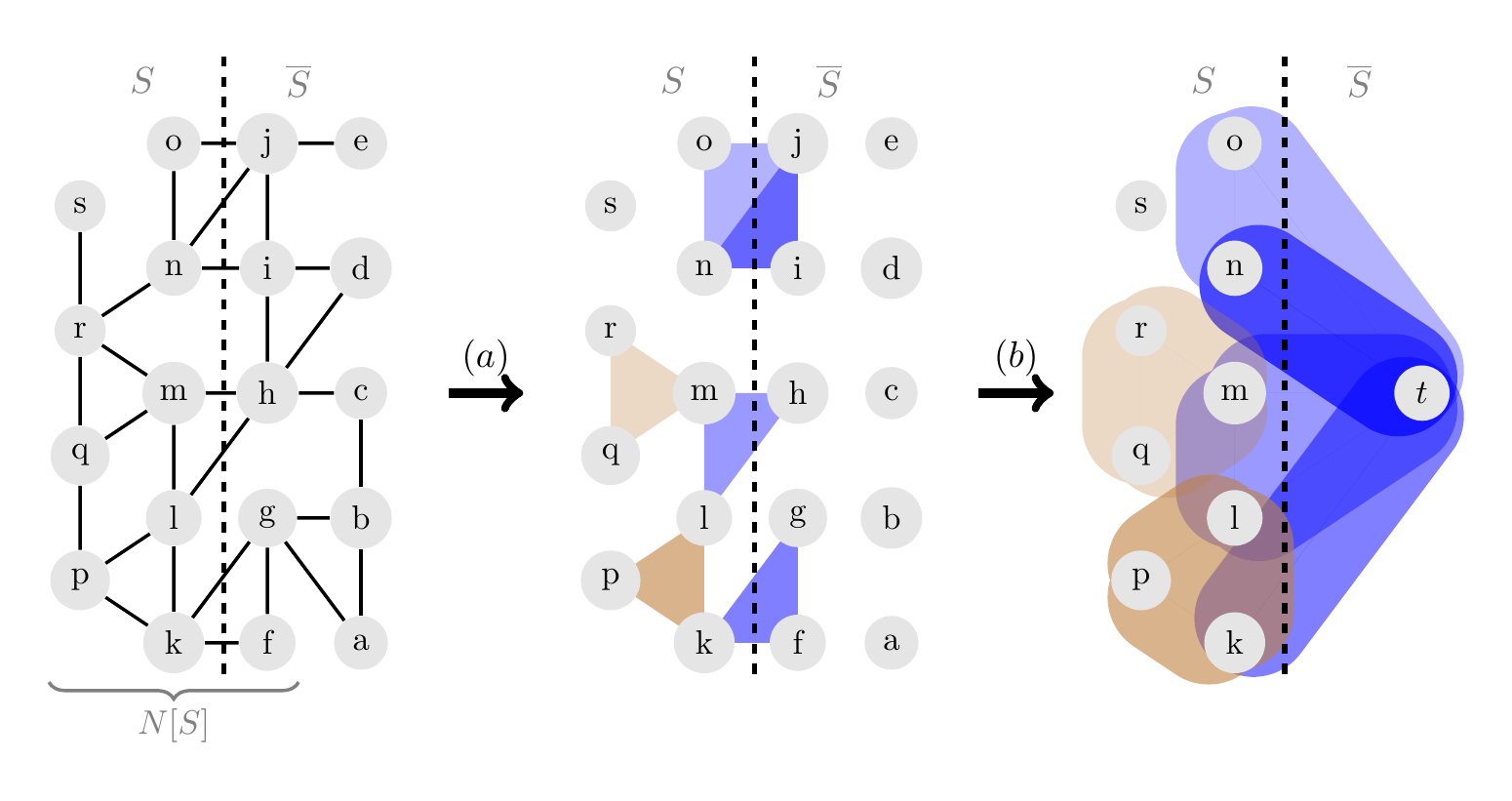}
	\caption{Example of motif-enumeration and model-construction phases of our algorithm for triangle motif and the hypergraph model. In the left, the nodes of $G$ are split into sets $S$ and $\overline{S}$. In the center, motif occurrences containing nodes in $S$ are enumerated. In the right,~$H_\mu$ is built by converting motifs into nets and contracting $\overline{S}$ into a node~$t$.}
	\label{fig:model_construction}
\end{figure*}

Our hypergraph model~$H_\mu$ is built in two conceptual operations.
First, define a hypergraph containing $V$ as nodes and a set~$\mathcal{E}$ of nets such that, for each motif in~$M$, $\mathcal{E}$ has a net with pins equal to the endpoints of this motif. 
Then, we contract together all nodes in~$\overline{S}$ into a single node~$t$ and substitute parallel nets by a single net whose weight is equal to the summed weights of the removed parallel nets.
More formally, we define the hypergraph version of our model as $H_\mu = (S \cup \{t\},\mathcal{E})$ where the set~$\mathcal{E}$ of nets contains one net~$e$ associated with each motif occurrence $G'=(V',E') \in M$ such that $e = V'$ if $V'\subseteq S$, and $e = V' \cap S \cup \{t\}$ otherwise.
In the former case the net has weight~$1$, in the latter case the net has weight equal to the amount of motif occurrences in $M$ represented by it.
\cschreplace{In a typical hypergraph contraction, the weight of $t$ would be $\mathfrc{c}(t) = c(\overline{S})$.
	Nevertheless, we opt to make the nodes of $H_\mu$ unweighted, which is convenient for the purpose of our overall algorithm as will become clear later.}
{The weight of $t$ is set to $\mathfrc{c}(t) = c(\overline{S})$. In our experiments, we set the nodes to unweighted for the partitioning process as the partitioners would have to deal with highly imbalanced nodes weights otherwise. Evaluating the objective is still done using the correct weights.}
In practice, the hypergraph version of $H_\mu$ can be built by instantiating the nodes in $S \cup \{t\}$ and the nets in $\mathcal{E}$.
Assuming that the number of nodes in $\mu$ is a constant, our model is built in time $O(|S|+|M|)$ and uses memory $O(|S|+|M|)$.
The construction of~$H_\mu$ is illustrated in transformation~(c) of Figure~\ref{fig:overall_algorithm} and demonstrated for a particular example in transformation~(b)~of~Figure~\ref{fig:model_construction}.

Observing the relationship between $G$ and $H_\mu$, we can distinguish three groups of components.
The first group comprises nodes in $S$ and motifs with all endpoints in $S$, all of which are represented in $H_\mu$ without any contraction as nodes and nets.
The second group consists of nodes in $\overline{S}$ and motifs with all endpoints in $\overline{S}$, which are compactly represented in $H_\mu$ as the contracted node~$t$.
The third group comprises motifs with nodes in both $S$ and $\overline{S}$, all of which are abstractly represented in $H_\mu$ as nets containing individual pins in $S$ as well as the pin~$t$.
Summing up, our hypergraph model is a concise representation of the whole graph $G$ where relevant information for local motif clustering is emphasized in two perspectives:
Edges are omitted while motifs are made explicit and global information is abstracted while local information is preserved in detail.
Theorem~\ref{theo:cut_equivalence} shows that the cut-net of a partition of $H_\mu$ directly corresponds to the motif-cut of an equivalent partition in $G$ if our motif enumeration step is exact.
On top of that, Theorem~\ref{theo:conductance_equivalence} shows that the motif conductance of this equivalent partition of $G$ can be directly computed from $H_\mu$ assuming $d_\mu(S) \leq d_\mu(\overline{S})$.
Assuming $d_\mu(S) \leq d_\mu(\overline{S})$ is fair since $S$ is ideally much smaller than~$\overline{S}$. 
Enumerating the motifs in $\overline{S}$ is not reasonable for a local clustering algorithm, but we did verify that our assumption holds during all our experiments.

\begin{mytheorem}
	Any $k$-way partition $P$ of our hypergraph model $H_\mu$ corresponds to a unique $k$-way partition $P^\prime$ of $G$ such that the cut-net of $P$ is equal to the motif-cut of $P^\prime$, assumed an exact motif enumeration step.
	\label{theo:cut_equivalence}
\end{mytheorem}

\begin{proof}
	For simplicity, we prove the claim assuming that parallel nets are not substituted by a single net whose weight is equal to their summed weights.
	This proof directly extends to our model since the contribution of a contracted cut net to the overall cut-net equals the contribution of the parallel nets represented by it.
	Due to the design of our hypergraph model $H_\mu$, there is a direct correspondence between its nodes and the nodes of $G$.
	Hence, any partition $P$ of $H_\mu$ corresponds to a partition $P^\prime$ of $G$ where corresponding nodes are simply assigned to the same blocks.
	Since $\overline{S}$ is represented by the single node~$t$ in $H_\mu$, no motif occurrence totally contained in $\overline{S}$ can be cut in $P^\prime$.
	All the remaining motif occurrences in $G$ can be potentially cut in $P^\prime$, but these motif occurrences are bijectively associated with the nets of $H_\mu$ with a direct correspondence between motif endpoints in $G$ and net pins in $H_\mu$.
	As a consequence, a motif occurrence of $G$ is cut in $P^\prime$ if, and only if, the corresponding net in $H_\mu$ is cut in~$P$.
\end{proof}

\begin{mytheorem}
	Given a $2$-way partition $P=(C,\overline{C})$ of our hypergraph model $H_\mu$ with~$t \in \overline{C}$, the motif conductance $\phi_\mu(C^\prime)$ of the corresponding $2$-way partition $P^\prime=(C^\prime,\overline{C^\prime})$ of $G$ is the ratio of the cut-net of~$P$ to $\mathcal{d}_{\mathfrc{w}}(C)$, assumed an exact motif enumeration step and $d_\mu(S) \leq d_\mu(\overline{S})$.
	\label{theo:conductance_equivalence}
\end{mytheorem}

\begin{proof}
	From Theorem~\ref{theo:cut_equivalence}, the motif-cut of $P^\prime$ can be substituted by the cut-net of $P$ in the numerator of the definition of $\phi_\mu(C^\prime)$.  
	To complete the proof, it suffices to show that the denominator of $\phi_\mu(C^\prime)$, namely $min(d_\mu(C^\prime),d_\mu(\overline{C^\prime}))$, is equal to $\mathcal{d}_{\mathfrc{w}}(C)$.
	Due to the design of $H_\mu$, the values of $d_\mu(C^\prime)$ and $\mathcal{d}_{\mathfrc{w}}(C)$ are identical.
	Our assumption~$t \in \overline{C}$ leads to $\overline{S} \subseteq \overline{C^\prime}$ and $C^\prime \subseteq S$, which respectively imply $d_\mu(\overline{S}) \leq d_\mu(\overline{C^\prime})$ and $d_\mu(C^\prime) \leq d_\mu(S)$. 
	Since $d_\mu(S) \leq d_\mu(\overline{S})$, hence $\mathcal{d}_{\mathfrc{w}}(C) = d_\mu(C^\prime) \leq d_\mu(S) \leq d_\mu(\overline{S}) \leq d_\mu(\overline{C^\prime})$.
\end{proof}

\subsection{Graph Model}
\label{subsec:Graph model}

In this section, we describe the graph version of our model~$H_\mu$ and explain how to build it.
Similarly to the hypergraph version of this model, our graph model can be built in time linear on the amount of nodes in~$S$ and motifs in~$M$.
We discuss advantages and limitations of the graph model in comparison to the hypergraph model.

\begin{figure*}[t!]
	\centering
	\includegraphics[width=.8\linewidth]{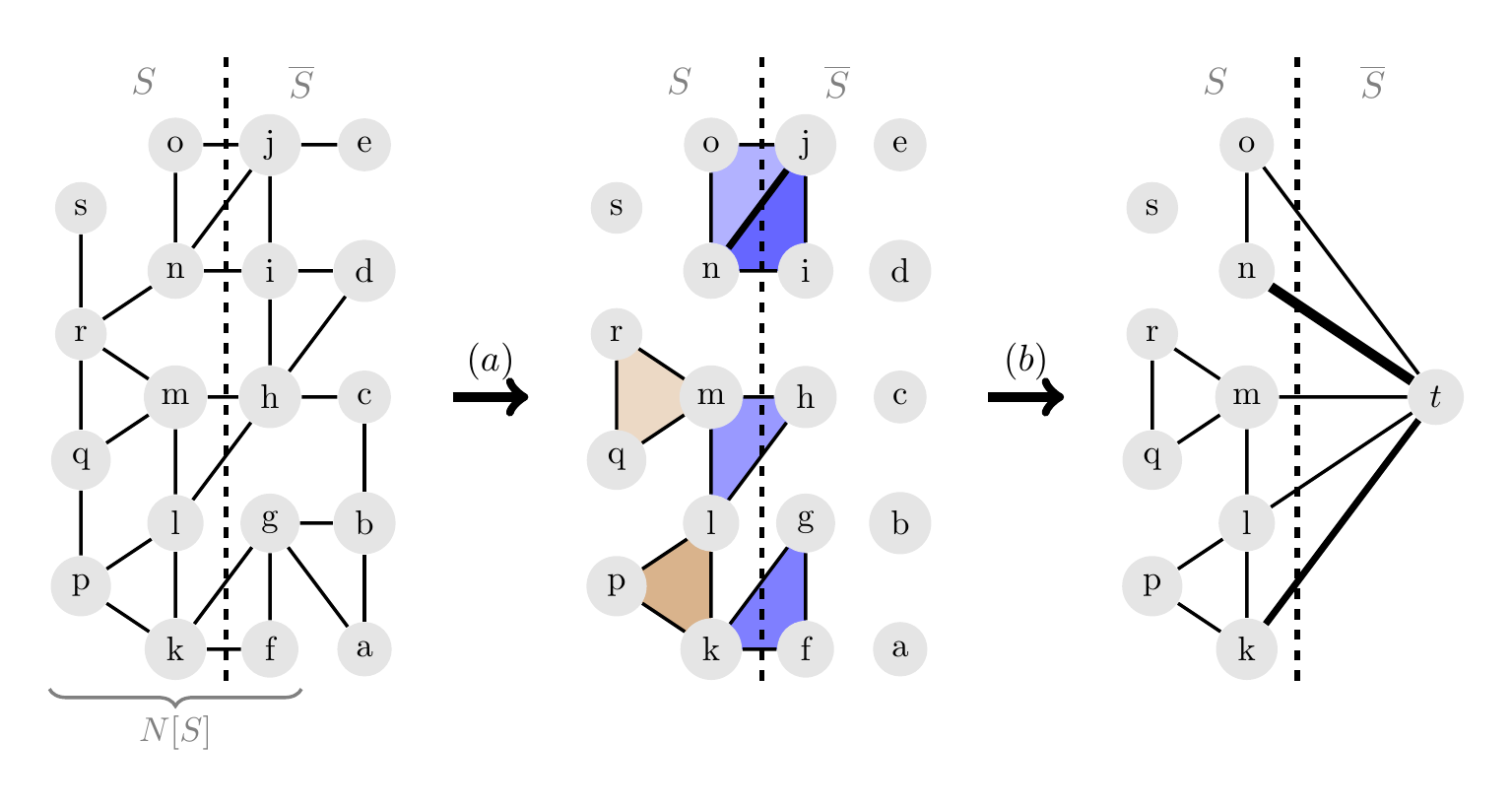}
	\caption{Example of motif-enumeration and model-construction phases of our algorithm for triangle motif and the graph model. In the left, $G$ is shown with its nodes split into the sets $S$ and $\overline{S}$. In the center, the motif occurrences containing at least a node in $S$ are enumerated and the weight of an edge equals the number triangles it touches. In the right, the model $H_\mu$ is built by contracting $\overline{S}$ into a single node~$t$. The weight of an edge is represented by its thickness.}
	\label{fig:model_construction_graph}
\end{figure*}

The graph version of our model~$H_\mu$ is built in two conceptual operations.
The first one consists of obtaining the weighted graph $W$ proposed by
\citet{benson2016higher} and used in the state-of-the-art algorithm MAPPR~\cite{yin2017local}.
The graph~$W$ contains $V$ as nodes and a set of edges such that two conditions are met: 
(i)~there is an edge between a pair of nodes if, and only if, both nodes belong at the same time to at least a motif in $G$; 
(ii)~the weight of an edge is equal to the number of motif occurrences containing both its endpoints.
The second operation consists of contracting all nodes in~$\overline{S}$ into a single node~$t$ and substitute parallel edges by a single edge whose weight is equal to the summed weights of the removed edges.
The construction of the graph version of~$H_\mu$ is illustrated in transformation~(c) of Figure~\ref{fig:overall_algorithm} and demonstrated for a particular example in transformation~(b) of Figure~\ref{fig:model_construction_graph}.
More formally, we define the model as $H_\mu=(S \cup \{t\},{E}_\mu)$ where~${E}_\mu$ contains an edge~$e$ for each pair of nodes sharing a motif $G'=(V',E') \in M$ provided that at least one of its endpoints is contained in $S$. 
The weight of $t$ is set to ${c}(t) = c(\overline{S})$.
Similarly to our approach with the hypergraph version of the model, we opt to make the nodes of the graph model unweighted in our experiments. 
In practice, the graph version of $H_\mu$ can be built by instantiating the nodes in $S \cup \{t\}$ and directly the computing the edges in ${E}_\mu$ and their weights.
Assuming that the number of nodes in $\mu$ is a constant, our graph model is built in time $O(|S|+|M|)$ and uses memory $O(|S|+|E_\mu|)$.
Especially for the triangle motif, the memory requirement of the graph model is $O(|S|+|N[S] \times N[S] \cap E|)$, which is linear on $n$ and $m$ in the worst~case.

We reproduce here Theorem~\ref{theo:conductance_equivalence_graph_complete} by \citet{yin2017local}, which shows that conductance in the weighted graph $W$ is equivalent to motif conductance in $G$ as long as the motif has at most $3$ nodes.
Based on this result, Theorem~\ref{theo:conductance_equivalence_graph} shows that we can compute motif conductance directly from our graph model $H_\mu$ if $d_\mu(S) \leq d_\mu(\overline{S})$.
As we mentioned, assuming $d_\mu(S) \leq d_\mu(\overline{S})$ is fair since $S$ is ideally much smaller than~$\overline{S}$. 
Recall that the hypergraph version of our model~$H_\mu$ is flexible enough to represent any motif as a net such that Theorems~\ref{theo:cut_equivalence}~and~\ref{theo:conductance_equivalence} continue valid.
Although the graph version of our model can technically represent any motif, Theorem~\ref{theo:conductance_equivalence_graph} is only valid for motifs with at most three nodes, while other models only allow a heuristic computation of the motif conductance~\cite{benson2016higher}.
Nevertheless, the biggest drawback of the hypergraph-based approach is the need for storing up to $|M|$ nets, which costs $O(n^{3})$ in the worst case.
In contrast, the memory needed to store our graph model is $O(n^2)$ in the worst case and $O(n+m)$ specifically for the triangle motif.

\begin{mytheorem}[Theorem 4.1 by \citet{yin2017local}]
	Given a $2$-way partition $P^{\prime\prime}=(C^{\prime\prime},\overline{C^{\prime\prime}})$ of the weighted graph $W$, the motif conductance $\phi_\mu(C^\prime)$ of the corresponding $2$-way partition $P^\prime=(C^\prime,\overline{C^\prime})$ in $G$ is equal to the conductance $\phi(C^{\prime\prime})$ of $C^{\prime\prime}$ in $W$ assumed a motif with at most three nodes.
	\label{theo:conductance_equivalence_graph_complete}
\end{mytheorem}

\begin{mytheorem}
	Given a $2$-way partition $P=(C,\overline{C})$ of the graph version of model $H_\mu$ with~$t \in \overline{C}$, the motif conductance $\phi_\mu(C^\prime)$ of the corresponding $2$-way partition $P^\prime=(C^\prime,\overline{C^\prime})$ of $G$ is the ratio of the edge-cut of~$P$ to ${d}_\omega(C)$, assumed an exact motif enumeration step, $d_\mu(S) \leq d_{\mu}(\overline{S})$, and a motif with at most three nodes.
	\label{theo:conductance_equivalence_graph}
\end{mytheorem}

\begin{proof}
	From Theorem~\ref{theo:conductance_equivalence_graph_complete}, the motif conductance of any $2$-way partition of $G$ is equal to the conductance of the equivalent partition in $W$ assuming a motif with at most $3$ nodes. 
	Since we assume $d_\mu(S) \leq d_\mu(\overline{S})$, hence the conductance in $W$ of any community $C^{\prime\prime} \in S$ is equal to its edge-cut divided by the volume of $C^{\prime\prime}$.
	From the construction of our graph model $H_\mu$, the assumed community~$C$ has an equivalent community~$C^{\prime\prime}$ in $W$ with same edge-cut and same volume, which completes the proof.
\end{proof}

\subsection{Partitioning}
\label{subsec:Partitioning}

In this section, we describe the (hyper)graph partitioning phase of our local motif clustering algorithm.
We present the used (hyper)graph partitioning algorithms and discuss how we enforce feasibility of the found solution and maximize its quality.
Moreover, we provide remarks about the running time of our partitioning phase.

The partitioning phase of our algorithm consists of a $2$-way partitioning of $H_\mu$. 
When using the hypergraph model, the partition is computed by the multi-level hypergraph partitioner KaHyPar~\cite{schlag2016k}. 
When using the graph model, the partition is computed by the multi-level graph partitioner KaHIP~\cite{kaHIPHomePage}.
These partitioners contain sophisticated algorithms to produce low-cut partitions of (hyper)graphs efficiently.
As already shown,
any $2$-way partition of $H_\mu$ automatically corresponds to a community in $G$.
Nevertheless, our aim is to obtain a \emph{consistent} partition of $H_\mu$, which we define as a partition where the seed node~$u$ and the contracted node~$t$ are in different blocks.
This consistent partition ultimately corresponds to a local community in~$G$ which contains the seed node~$u$ and is completely contained in the ball~$S$.
This \emph{consistency} criterion is important since the nodes in $\overline{S}$ have not been explored by our algorithm and are farther from $u$ than the nodes in~$S$.
Note that the $2$-way partition illustrated in Figure~\ref{fig:overall_algorithm} generates a consistent local community according to our definition of it.
While KaHyPar allows partitioning hypergraphs with fixed nodes, KaHIP does not offer such functionality.
Nevertheless, we ensure block feasibility for both versions of our algorithm by simply assigning the seed node to the block that does not contain $t$ after the partition is computed (before computing the motif conductance).
Although simplistic, this approach has not affected solution quality considerably for the hypergraph-based version of our algorithm in preliminary comparisons against a fixed nodes-based approach.

Besides corresponding to a consistent local clustering on the original graph, our solution should have as low a motif conductance as possible.
However, KaHyPar and KaHIP \addcsch{are randomized algorithms and} do not directly optimize for this objective, but rather minimize the cut value while enforcing a hard balancing constraint.
To improve our results, we explore different combinations of edge-cut (resp. cut-net) and imbalance by repeating the partitioning procedure $\beta$~times with random balancing constraints for each built (hyper)graph model, where $\beta$ is a tuning parameter.
For each obtained partition, our algorithm computes the motif conductance of the corresponding local cluster as shown in Theorem~\ref{theo:conductance_equivalence_graph} (resp. Theorem~\ref{theo:conductance_equivalence}) and keeps the partition associated with the lowest motif~conductance.

KaHyPar and KaHIP run in time close to linear in practice.
Nevertheless, the partitioning phase can be the dominating operation of our overall local clustering algorithm. 
This is the case because KaHyPar and KaHIP use sophisticated algorithms and data structures in order to minimize the cut value, which increases constant factors in the algorithm running time complexity.
KaHyPar can be especially much slower than KaHIP since the number of nets in the hypergraph model can be considerably larger than the number of edges in the graph model.
\cschreplace{We can make the hypergraph version of our algorithm faster by using Mt-KaHyPar~\cite{gottesburen2021scalable} instead of KaHyPar.
	Mt-KaHyPar obtains significant speedups by using allowing shared-memory parallel execution.}{In both cases, running time can be improved further by using parallelized tools such as Mt-KaHyPar~\cite{gottesburen2021scalable} or KaMinPar~\cite{DBLP:conf/esa/GottesburenH00S21}. However, parallelization is not the focus~of~this~work.}

\textbf{Local Search.}
\label{subsec:Local Search}
We implement a local search inspired by \emph{label propagation} \cite{labelpropagationclustering} for the graph model-based version of our algorithm. 
This local search is designed to optimize for the correct objective, i.e., motif conductance.
We apply it directly on each $2$-way partition generated by KaHIP in order to increase the chances of reaching a local minimum motif conductance.
The local search algorithm works in rounds.
In each round, it visits all nodes of $H_\mu$ in a random order, starting with the labels being the current assignment of nodes to blocks.
When a node $v$ is visited, it is moved to the opposite block if this movement causes a decrease in the motif conductance of the clustering.
Movements of nodes with zero-gain can occasionally occur with $50\%$ probability.
We ensure that the seed node $u$ and the contracted node $t$ continue in opposite blocks by simply skipping them.
We stop the local search when a local optimum is reached or after at most~$\ell$ rounds, where $\ell$ is a tuning parameter.

\section{Experimental Evaluation}
\label{sec:Experimental Evaluation}

\paragraph*{Methodology.} 
We performed the implementation of our algorithm in C++ on the KaHIP framework using the public libraries for KaHyPar~\cite{schlag2016k} and KaHIP~\cite{kaffpa}. 
We use the fastest configuration of these tools throughout our experiments.
\addcsch{Note that using stronger configurations would likely yield better solutions at the cost of higher running time.}
We compiled our program using gcc 9.3 with full optimization turned on (-O3 flag).
For the reported experiments, we use a time limit of one hour for our overall algorithm.
This time limit is checked between repetitions of the partitioning phase, hence it can be violated in case a particular partitioning procedure takes too long.
All our experiments are based on the triangle motif, i.e., the undirected clique of size $3$.
We ensure the integrity of our results by using the same motif-conductance evaluator function for all tested algorithms.
In our experiments, we have used a machine with a sixty-four-core AMD EPYC 7702P processor running at $2.0$ GHz, $1$ TB of main memory, $32$ MB of L2-Cache, and $256$ MB of L3-Cache. 
We measure running time, motif-conductance, and/or size of the computed cluster.
For each graph, we pick $50$ random seed nodes and use all of them as input for each algorithm.
When averaging running time or cluster size over multiple instances, we use the geometric mean in order to give every instance the same influence on the \textit{final score}. 
When averaging motif conductance over multiple instances, the final score is computed via arithmetic mean.
This is a necessary averaging strategy since motif conductance can be zero, which makes the geometric mean infeasible to compute.
We also use \emph{performance profiles} which relate the running time (resp. motif conductance) of a group of algorithms to the fastest (resp. best) one on a per-instance basis.
Their x-axis shows a factor~$\tau$ while their y-axis shows the percentage of instances for which algorithm $A$ has up to~$\tau$ times the running time (resp. motif conductance) of the fastest (resp. best) algorithm.

\paragraph*{Instances.}
We use graphs from various sources \cite{snapnets,nr-aaai15,benchmarksfornetworksanalysis} to test our algorithm.
Most of the considered graphs were used for benchmark in previous works in the area.
Prior to our experiments, we removed parallel edges, self-loops, and directions of edges and assigning unitary weight to all nodes and edges.
Basic properties of the graphs under consideration can be found in Table~\ref{tab:graphs}.
For our experiments, we split the graphs in two disjoint sets: a \emph{tuning} set for the parameter study experiments and a \emph{test} set for the comparisons against the state-of-the-art. The graphs in the test set are exactly the graphs used in the MAPPR paper \cite{yin2017local}.

\begin{table*}[]
	\centering
	\setlength{\tabcolsep}{28pt}
	\begin{tabular}{| l  r  r  r | }
		\hline
		Graph & $n$& $m$ & Triangles\\
		\hline  \hline

		\multicolumn{4}{|c|}{Tuning Set} \\
		\hline

		citationCiteseer & \numprint{268495}    & \numprint{1156647}  & \numprint{847420} \\
		
		coAuthorsCiteseer & \numprint{227320}    & \numprint{814134}  & \numprint{2713298} \\
		
		amazon0312 & \numprint{400727}  & \numprint{2349869} & \numprint{3686467} \\
		
		amazon0505 & \numprint{410236} & \numprint{2439437} & \numprint{3951063} \\
		
		amazon0601 & \numprint{403364} & \numprint{2443311} & \numprint{3986507} \\
		
		del22 & \numprint{4194304}  & \numprint{12582869} & \numprint{8436672} \\
		
		del23 & \numprint{8388608}  & \numprint{25165784} & \numprint{16873359} \\
		
		soc-pokec & \numprint{1632803} & \numprint{22301964}   & \numprint{32557458} \\
		
		rgg22 & \numprint{4194304} & \numprint{30359198} & \numprint{85962754} \\
		
		rgg23 & \numprint{8388608} & \numprint{63501393} & \numprint{188022664} \\
		
		in-2004 & \numprint{1382908}    & \numprint{13591473}  & \numprint{464257245} \\
		
		\hline

		\multicolumn{4}{|c|}{Test Set} \\
		\hline

		com-amazon & \numprint{334863}  & \numprint{925872} & \numprint{667129} \\
		
		com-dblp & \numprint{317080}  & \numprint{1049866} & \numprint{2224385} \\
		
		com-youtube & \numprint{1134890}  & \numprint{2987624} & \numprint{3056386} \\
		
		com-livejournal & \numprint{3997962}  & \numprint{34681189} & \numprint{177820130} \\
		
		com-orkut & \numprint{3072441}  & \numprint{117185083} & \numprint{627584181} \\
		
		com-friendster & \numprint{65608366} & \numprint{1806067135} & \numprint{4173724142} \\		
		\hline
		
	\end{tabular}
	\caption{Graphs for experiments.}
	
	\label{tab:graphs}
\end{table*}

\paragraph{Parameter Study.}
\label{subsec:Parameter Study}

We performed extensive tuning experiments using the graphs disjoint from the graphs used for the evaluation against state-of-the-art. 
Due to space constraints, we only summarize the main results.
In a comparison of the hypergraph-based version of our algorithm against its graph-based version, each approach produces the best motif conductance for around $50\%$ of the instances.
Nevertheless, the hypergraph-based version is $23$ times slower on average and uses up to $68.7$ times more memory since the hypergraph model stores a large number of nets.
Hence, we exclusively use the graph-based version of our algorithm for the remaining experiments.
Next, we compare the effect of using different values for the parameters $\beta$ and $\alpha$.
The results show the expected regular relationship between solution quality, running time, and these parameter: the larger $\alpha$ (resp. $\beta$), the smaller the motif conductance and the larger the running time.
Finally, we checked that the impact of including the label propagation local search in our algorithm is, on average, a $13\%$ decrease in the motif conductance at the cost of only $1.5\%$ more running time.

\subsection{Comparison against State-of-the-Art}
\label{subsec:Comparison against State-of=the-Art}

In this section, we show experiments in which we compare our algorithm against the state-of-the-art, namely MAPPR~\cite{yin2017local}.
We were not able to compare against other algorithms for one or both of the following reasons: 
(i)~code is not available~\cite{rohe2013blessing,zhang2019local,meng2019local,shang2022local},
(ii)~algorithm solves a different problem, or
optimizes for a different objective~\cite{huang2014querying}.
An exception for these reasons is HOSPLOC~\cite{zhou2021high}.
The algorithm is implemented in Python and very slow even for small graphs. 
Moreover, experiments done in their paper are on graphs that are  multiple orders of magnitude smaller than the graphs used in our evaluation. 
Hence, we are not able to perform comparisons against it.
We compare our results against the globally best cluster computed by MAPPR for each seed node using standard parameters ($\alpha=0.98$, $\epsilon=10^{-4}$).
Unless mentioned otherwise, experiments presented here involve all graphs from the Test Set in Table~\ref{tab:graphs}.

\begin{table*}[b!]
	\centering
	\setlength{\tabcolsep}{16pt}
	\begin{tabular}{|l|rrr|rrr|}
		\hline
		\multirow{2}{*}{Graph} & \multicolumn{3}{c|}{GL;3;80}    & \multicolumn{3}{c|}{MAPPR} \\
		& $\phi_\mu$     & $|C|$   & t(s)   & $\phi_\mu$  & $|C|$ & t(s)    \\ \hline \hline
		com-amazon             & \textbf{0.037} & 64    & 0.22  & 0.153      & 58  & 2.68   \\
		com-dblp               & \textbf{0.115} & 56    & 0.38  & 0.289      & 35  & 3.04   \\
		com-youtube            & \textbf{0.172} & 1443  & 7.93  & 0.910      & 2   & 10.44  \\
		com-livejournal        & \textbf{0.244} & 387   & 8.17  & 0.507      & 61  & 173.80 \\
		com-orkut              & \textbf{0.150} & 13168 & 496.94 & 0.407      & 511 & 923.26 \\
		com-friendster         & \textbf{0.368}      &  10610     &   1339.99     &   0.741         &   121  &   16565.99      \\ 
		\hline
		Overall                & \textbf{0.181} & 823   & 12.67  & 0.500      & 50  & 79.34  \\ \hline
	\end{tabular}
	\caption{Average comparison against state-of-the-art.}
	
	\label{tab:resultsoverall}
\end{table*}

\begin{figure}[t]{}
	\centering
	\includegraphics[width=0.525\textwidth]{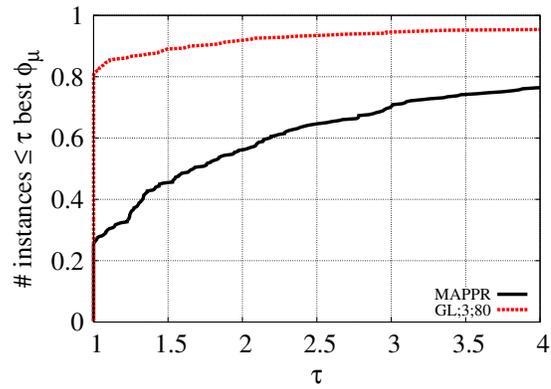}
	\vspace*{-0.5cm}
	\caption{Motif conductance performance profile.}
	\vspace*{-0.5cm}
	\label{fig:SIMPLEstateoftheart_graph_res_pp}
\end{figure}

\begin{figure}[t]{}
	\centering
	\includegraphics[width=0.525\textwidth]{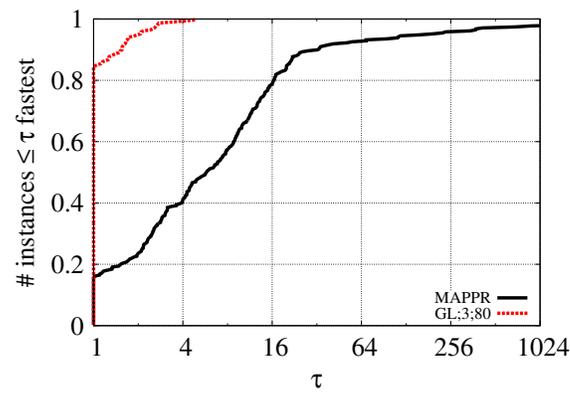}
	\vspace*{-0.5cm}
	\caption{Running time performance profile.}
	\vspace*{-0.5cm}
	\label{fig:SIMPLEstateoftheart_graph_tim_pp}
\end{figure}

\begin{figure}[h!]{}
	\centering
	\includegraphics[width=0.525\textwidth]{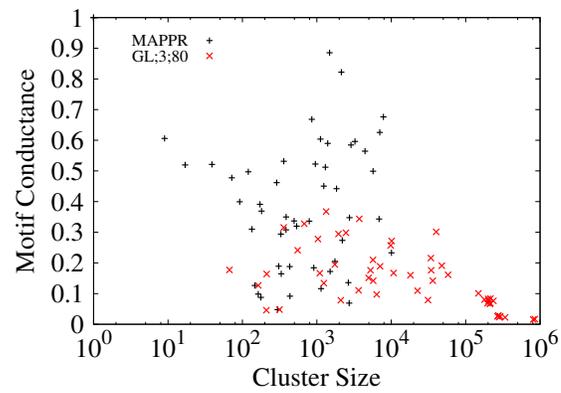}
	\vspace*{-0.5cm}
	\caption{Motif conductance vs cluster size for com-orkut.}
	\label{fig:SIMPLEstateoftheart_graph_res_double}
\end{figure}

In the performance profile plots shown in Figures~\ref{fig:SIMPLEstateoftheart_graph_res_pp}~and~\ref{fig:SIMPLEstateoftheart_graph_tim_pp}, we compare MAPPR against our algorithm.
More specifically, we use the following parameters for our algorithm: graph-based model, $\alpha=3$, $\beta=80$, and label propagation local search.
Our algorithm obtains the best \addcsch{or equal} motif conductance \addcsch{value} for around $80\%$ of the instances, while MAPPR obtains the best \addcsch{or equal} result for around $25\%$ of the instances.
This result can be explained by the fact that our algorithm explores the solution space better than MAPPR, since we perform multiple cluster constructions and refinements, while MAPPR simply uses the  APPR algorithm.
At the same time, note that our algorithm is faster than MAPPR for $84\%$ of the instances, besides being on average a factor $6.3$ faster (in our experiments up to $3$ orders of magnitude).
The main explanation for this considerable running time difference is the fact that MAPPR has to enumerate motifs throughout the whole graph in order to operate, whereas our algorithm only needs to enumerate motifs in a ball of nodes around the seed node.
In Table~\ref{tab:resultsoverall}, we show average results for each graph in our Test Set as well as average results overall.
Note that on average our algorithm outperforms MAPPR with respect to motif conductance and running time for every single graph and overall.
\addcsch{On average our algorithm computes a motif conductance value of 0.181 while MAPPR computes an average value of 0.500.}
Finally, Figure~\ref{fig:SIMPLEstateoftheart_graph_res_double} plots motif conductance vs cluster size for the communities computed for the com-orkut graph.
Note that the communities found by our algorithm are visibly localized in the lowest half of the chart, while the communities computed by MAPPR are~more~widespread.

\section{Conclusion}
\label{sec:Conclusion}

We proposed an algorithm which computes local motif clustering via partitioning of (hyper)graph models. 
Given a seed node, our algorithm selects a ball of nodes around it and builds a (hyper)graph model which is designed such that an optimal solution in the (hyper)graph model minimizes the motif conductance in the original network.
In extensive experiments with the triangle motif, we observe that our algorithm computes communities that have on average one third of the motif conductance value than MAPPR while being $6.3$ times faster on average and removing the necessity of a preprocessing motif-enumeration on the whole network.

\begin{acks}
 Partially supported by DFG grant SCHU 2567/1-2.
\end{acks}

\bibliographystyle{ACM-Reference-Format}
\bibliography{phdthesiscs.bib}

\end{document}